\documentclass[acmsmall]{acmart}
\def\FIGDIR{./figures}          

\usepackage{algorithmic}
\usepackage[linesnumbered]{algorithm2e}

\usepackage{makecell}
\usepackage{tabu}
\usepackage{setspace}
\usepackage{comment}
\usepackage[most]{tcolorbox}
\usepackage{pifont}

\usepackage{tabularx}

\usepackage{endnotes,xspace,graphicx,fancyvrb,multirow}

\usepackage{paralist}

\usepackage{fancyhdr}
\usepackage[normalem]{ulem}

\usepackage{booktabs} 

\usepackage{enumitem}
\usepackage{enumerate}

\usepackage{caption}
\usepackage[keeplastbox]{flushend}

\RequirePackage[nameinlink]{cleveref} 
\crefname{chapter}{Chapter}{Chapters}
\crefname{section}{Section}{Sections}
\crefname{subsection}{Section}{Sections}
\crefname{equation}{Equation}{Equations}
\crefname{definition}{Definition}{Definitions}
\crefname{assumption}{Assumption}{Assumptions}
\crefname{theorem}{Theorem}{Theorems}
\crefname{figure}{Fig.}{Figures}
\crefname{table}{Table}{Tables}
\let\autoref\cref 

\newcommand{\PCignore}[1]{}

\newcommand{\mycaption}[1]{
  \caption{\textbf{#1}}
}

\newcommand{\insertFigureWidth}[3]{
    \begin{figure}[t]
        \centering
        \includegraphics[width=#2\linewidth]{\FIGDIR/#1.pdf}
        \mycaption{#3}
        \label{fig:#1}
    \end{figure}
}

\newcommand{\insertFigureWidthBump}[4]{
    \begin{figure}[t]
        \centering
        \includegraphics[width=#2\linewidth]{\FIGDIR/#1.pdf}
 	\vspace{#3}
        \mycaption{#4}
        \label{fig:#1}
    \end{figure}
}

\newcommand{\insertWideFigure}[2]{

    \begin{figure*}[t]
        \centering
        \includegraphics[width=\textwidth]{\FIGDIR/#1.pdf}
        \mycaption{#2}
        \label{fig:#1}
    \end{figure*}
}

\newcommand{\squishlist}{
 \begin{list}{$\bullet$}
  { \setlength{\itemsep}{0pt}
     \setlength{\parsep}{3pt}
     \setlength{\topsep}{3pt}
     \setlength{\partopsep}{0pt}
     \setlength{\leftmargin}{1.5em}
     \setlength{\labelwidth}{1em}
     \setlength{\labelsep}{0.5em} } }

\newcommand{\squishlisttwo}{
 \begin{list}{$\bullet$}
  { \setlength{\itemsep}{0pt}
     \setlength{\parsep}{0pt}
    \setlength{\topsep}{0pt}
    \setlength{\partopsep}{0pt}
    \setlength{\leftmargin}{2em}
    \setlength{\labelwidth}{1.5em}
    \setlength{\labelsep}{0.5em} } }

\newcommand{\squishend}{
  \end{list}  }

\newcommand{\betterparagraph}[1]{\noindent\textbf{#1.}}

\newcommand{\TODO}[1]{\textcolor{red}{TODO: #1}}
\newcommand{\HK}[1]{\textcolor{OliveGreen}{HK: #1}}

\newcommand{\rev}[1]{\textcolor{black}{#1}}

\definecolor{amber}{rgb}{0.75,0.35,0.0}
\newcommand{\hl}[1]{\textcolor{black}{#1}}
\usepackage{mdframed}

\newmdtheoremenv{definition}{Definition}
\newmdtheoremenv{theorem}{Theorem}
\newmdtheoremenv{assumption}{Assumption}

\newcommand{\inflex}{InFlex\xspace}
\newcommand{\partflex}{PartFlex\xspace}
\newcommand{\fullflex}{FullFlex\xspace}

\AtBeginDocument{%
  \providecommand\BibTeX{{%
    \normalfont B\kern-0.5em{\scshape i\kern-0.25em b}\kern-0.8em\TeX}}}

\setcopyright{rightsretained}
\usepackage{etoolbox}
\makeatletter
\patchcmd{\maketitle}{\@copyrightpermission}{
\begin{minipage}{0.3\columnwidth}
\href{http://creativecommons.org/licenses/by/4.0/}{\includegraphics[width=0.90\textwidth]{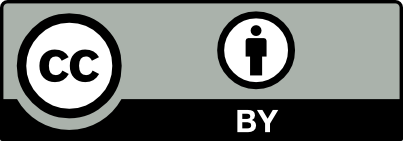}}
\end{minipage}\hfill
\begin{minipage}{0.7\columnwidth}
\href{http://creativecommons.org/licenses/by/4.0/}{This work is licensed under a Creative Commons Attribution International 4.0 License.}  
\end{minipage}}

\acmJournal{POMACS}
\acmYear{2022} \acmVolume{6} 
\acmNumber{2} \acmArticle{41} 
\acmMonth{6} \acmPrice{} 
\acmDOI{10.1145/3530907}

\makeatother

\received{February 2022}
\received[revised]{March 2022}
\received[accepted]{April 2022}

\begin{document}

\title{A Formalism of DNN Accelerator Flexibility}

\author{Sheng-Chun Kao}
\affiliation{%
  \institution{Georgia Institute of Technology}
  \country{USA}
  }
\email{felix@gatech.edu}

\author{Hyoukjun Kwon}
\affiliation{%
  \institution{Georgia Institute of Technology}
  \country{USA}
  }
\email{hyoukjun@gatech.edu}

\author{Michael Pellauer}
\affiliation{%
  \institution{NVIDIA}
  \country{USA}
  }
\email{mpellauer@nvidia.com}

\author{Angshuman Parashar}
\affiliation{%
  \institution{NVIDIA}
  \country{USA}
  }
\email{aparashar@nvidia.com}

\author{Tushar Krishna}
\affiliation{%
  \institution{Georgia Institute of Technology}
  \country{USA}
  }
\email{tushar@ece.gatech.edu}



\begin{abstract}

The high efficiency of domain-specific hardware accelerators for machine learning (ML) has come from \textit{specialization}, with the trade-off of less configurability/ flexibility. 
There is growing interest in developing \textit{flexible} ML accelerators to make them future-proof to the rapid evolution of Deep Neural Networks (DNNs). However, the notion of accelerator flexibility has always been used in an informal manner, restricting computer architects from conducting systematic apples-to-apples design-space exploration (DSE) across trillions of choices. In this work, we formally define accelerator flexibility and show how it can be integrated for DSE. 

Specifically, we capture 
DNN accelerator flexibility across four axes:
tiling, ordering, parallelization, and array shape. We categorize existing accelerators into 16 classes based on their axes of flexibility support, and define a precise quantification of the degree of flexibility of an accelerator across each axis. We leverage these to develop a novel flexibility-aware DSE framework. 
%
We demonstrate how this can be used to perform first-of-their-kind evaluations, including an isolation study to identify the individual impact of the flexibility axes. We demonstrate that adding flexibility features to a hypothetical DNN accelerator designed in 2014 improves runtime on future (i.e., present-day) DNNs by $11.8 \times$ geomean.


\end{abstract}



\maketitle

\section{Introduction}
\label{sec:introduction}


Machine learning (ML), especially deep learning (DL), has demonstrated great results in computer vision, natural language processing, and recommendation systems~\cite{Karras2019stylegan2, imagegpt, switchtransformer, dlrm19}. This has energized the field of computer architecture to develop customized hardware accelerator ASICs to enable deployment \rev{(i.e., inference)} of DL solutions in disparate environments like the edge and cloud. \rev{Once deployed, 
the overall runtime and energy-efficiency of running a DNN on the accelerator depends on its ~\emph{mapping}~\cite{kwon2020maestro, parashar2019timeloop}, which determines how tiles of computation are staged on the accelerator across space and time. Traditionally, the efficiency of domain-specific accelerator ASICs has come from \emph{specialization}, i.e., the control path and datapath in the accelerator are tailored to the deep neural networks (DNNs) that are expected to run on the accelerator. In other words, the number of mappings that an accelerator can support (aka \emph{map-space}) is restricted.}

\rev{Specialization is unfortunately a double-edged sword. While high-specialization enables DNN accelerator ASICs to provide higher performance and better energy-efficiency than CPUs and GPUs, it restricts the accelerator from adapting to the growing diversity in DNN models today. This is becoming a key challenge, given 
how rapidly the field of ML is evolving, since it is highly probable that accelerator chips will be obsolete by the time they reach the market.}

%
To this end, there has been growing interest in developing \textit{flexible} DNN accelerators. In contrast with a fully-programmable (i.e., Turing Complete) processor such as a CPU/GPU, or a fully-reconfigurable circuit such as a FPGA, a flexible accelerator adds small-overhead-but-high impact reconfigurability. 
\rev{The flexibility manifests in enabling the DNN accelerator to support diverse mapping strategies (i.e., larger map-spaces) for the same DNN layer. Intuitively, this works since diverse DNN layers can be thought of as different shapes and sizes of a few key fundamental operators such as convolutions or matrix-multiplications; the ability to change between mappings at runtime after silicon deployment allows the compiler to map skewed dimensions of different operations over space or time to enhance utilization and hide memory access times behind computations.}

\rev{Flexibility provides a two-fold benefit. First, it allows the accelerator to 
better tailor itself to the diverse set of layer parameters within the current DNN being mapped~\cite{kwon2020maestro, parashar2019timeloop}, instead of targeting the average case. For example, the CONV2\_1 early layer in ResNet-50 has 56x56 activations with 64 channels, whereas the CONV5\_3 late layer in ResNet-50  has 7x7 activations with 2048 channels.
An accelerator that only supports parallelism on activations can result in severe under-utilization on the late layer, and vice versa for one that only supports parallelism on channels. Second, it helps ``future-proof" the accelerator. For instance, consider an accelerator such as Eyeriss~\cite{chen2016eyeriss_jssc}, developed in 2014 for accelerating AlexNet~\cite{alexnet}, the state-of-the-art network at that time for image classification, which had 3x3, 5x5 and 11x11 filters; Eyeriss is severely under-utilized for newer DNN models that use layers such as pointwise (i.e., 1x1 filters) convolutions (e.g., MobileNetV2~\cite{sandler2018mobilenetv2}), as demonstrated by its successor Eyeriss\_v2~\cite{eyev2}, or attention (e.g., BERT~\cite{bert}).}

While the notion of flexibility described above makes intuitive sense, 
the field of DNN accelerators today is inconsistent
about the definition. We find that ``flexibility'' has been used loosely to refer either to the ability to handle 
different loop tiling/blocking~\cite{yang2020interstellar}, 
and/or support for different loop orders~\cite{lu2017flexflow}, 
and/or the ability to spatially partition across different dimensions~\cite{eyev2}, 
and/or the ability to 
logically 
support different PE aspect ratios~\cite{kwon2018maeri}. 
In fact, there is also inconsistency in whether flexibility is a hardware feature~\cite{kwon2020maestro} 
or simply a term for compiler loop-transformations over a fixed inner tile ~\cite{yang2020interstellar, sara_isca2021}.
This in-formalism is a serious barrier to the adoption of flexibility features, as it hinders precise quantification of the cost/benefit tradeoff.
Moreover, addition of large map-spaces confounds 
the accelerator 
Design-Space Exploration (DSE) problem
as concepts such as \textit{partial} flexibility cannot be precisely expressed or numerically assessed by existing DSE flows.

In this work, we present a novel formal definition and precise quantification of accelerator flexibility. \rev{We also quantify the corresponding cost for flexibility, including area and energy.} We then couple that with an abstraction that allows its incorporation into a first-of-its-kind \emph{flexibility-aware} automated DSE, without making the problem intractable. As a debut work in this field, we cannot claim to be exhaustive. However, we advocate that our formalism can be the first step toward the accelerator community moving towards consensus on the costs and benefits of flexible accelerators. Specifically, we make the following contributions:

\textbf{Contribution 1:} We present a formal definition of the flexibility of an accelerator that refers to the percentage of explorable / supported map-space out of all possible (i.e., exhaustive, as formally defined in Section \ref{sec:mapspace}) mappings for a DNN layer. We distill this down further into \textit{workload-dependent} (i.e., what a compiler/ mapping optimizer is legally allowed to explore) and a workload-independent but \textit{hardware-dependent} (i.e., what the hardware is designed to support) map-spaces.

\textbf{Contribution 2:} As these map-spaces can be overwhelmingly complex, we identify a complementary abstraction: we taxonomize flexibility into four axes -- tile size (T), order (O), parallelism (P) and array shape (S).
Identifying each axis with a binary 0/1 value
enables us to create 
16 flexibility classes within which 
various accelerators can fall, enabling a systematic classification of prior work.
For each axis, we identify both the hardware support needed, and the software loop-transformations it exposes to a mapper.

\textbf{Contribution 3:}
Together Contributions 1 and 2 allow us to precisely specify constraints on the shape of the map-space in order to distinguish diverse accelerators with \textit{varying degrees of flexibility} \rev{that we quantify with a new metric called
\emph{flexibility fraction} or \emph{flexion}\footnote{Pronounced ``fleck-shun.'' In biology this term refers to the act of bending a joint, making it apt for a flexibility metric.}} 
during automated DSE. 
We demonstrate flexibility-aware DSE using a genetic algorithm, constrained by the amount of flexibility provided by a target accelerator along each of the four axes. 

\textbf{Contribution 4:} 
We leverage the above contributions to perform isolation analysis, comparing several accelerator design points along each axis (independently and in conjunction), 
identifying the cost and benefit of flexibility for modern DNNs across vision, language and recommendation.
We also evaluate a first-of-its-kind ``what-if'' scenario to determine how a vision DNN accelerator designed in 2014 would have fared on DNN models today if it had been made future-proof by adding flexibility across one or more axes.
We summarize our key takeaways:
\squishlist

\item We find that making a 2014 DNN accelerator future-proof by increasing flexibility can significantly improve runtime on future (i.e., present-day) DNNs (11.8$\times$ geomean).

\item For Mnasnet, independently, (T) and (P) yield the most (4.8x/1.7x) performance gain and (S) the least (1.04x).
%

\item ``Partially'' flexible accelerators -- easier to build and program -- can sometimes provide similar benefits to fully-flexible ones (12-57\% depending on the network).


\item For DNN models formed with similar set of layers, (T) offered the most valuable flexibility (3.2x speedup). For diverse models, accelerators providing flexibility along (P) and (S) together offered higher benefits in runtime (49\%) compared to investing in only one of them (8-18\%).


\item Area overheads for flexibility are low ($<$1\%) in practice. Interestingly, we find no energy overhead for this area increase, as the features pay for themselves via better mappings that reduce DRAM accesses.

\squishend

In all, we hope that quantifying the benefits of flexibility can help the community better understand the value of investing engineering effort into flexible hardware accelerators, and the software mapper toolchains required to exploit it.

\vspace{-2mm}

\section{Background}
\label{sec:background}

\subsection{DNN Accelerator Components}

\insertFigureWidth{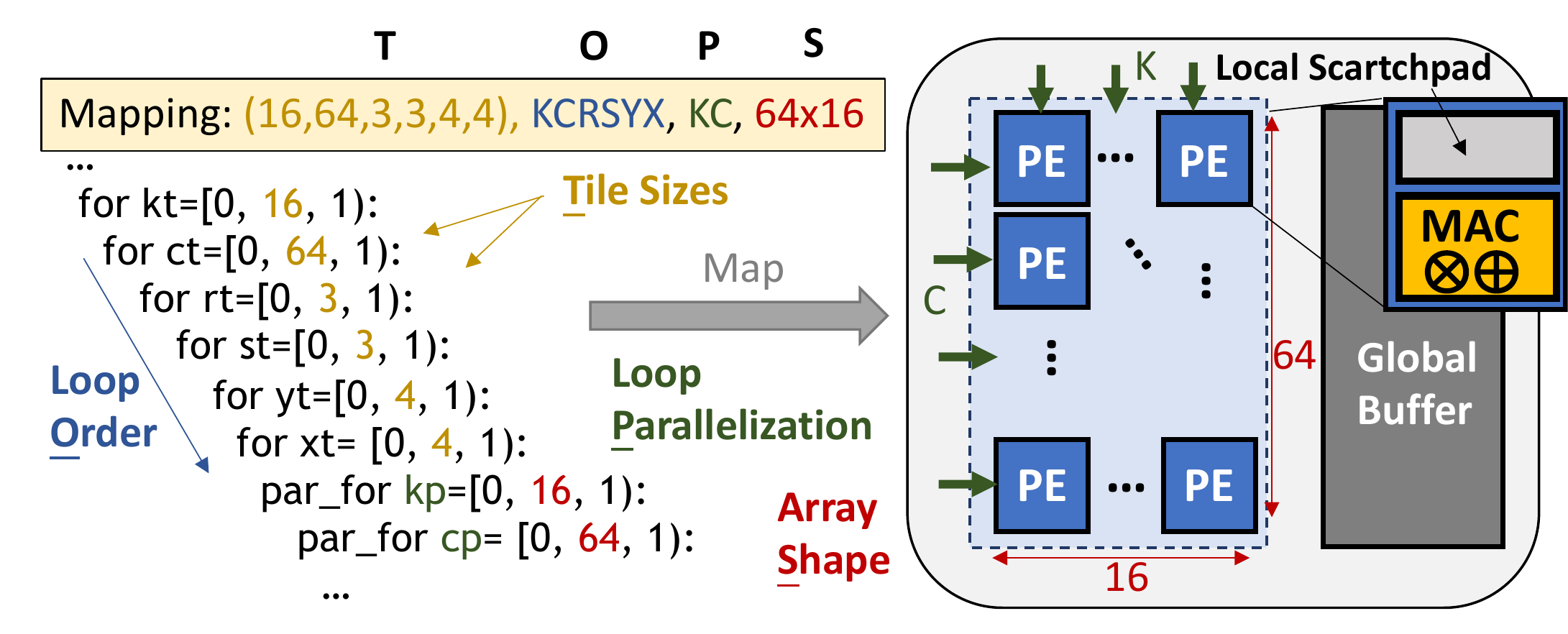}{0.75}{An example NVDLA mapping (K/C: Output/input channels. Y/X and R/S: height/width of activation and weight).}


\autoref{fig:nvdla} shows an example of 
a DNN accelerator. 
It is comprised of a spatial array of processing elements (PEs), buffers (across a memory hierarchy), and networks-on-chip (NoC) for operand distribution and output reduction/collection.
In \autoref{sec:hw_cost}, we discuss the hardware support needed in each of these axes to provide ``flexibility''.



\vspace{-1mm}
\subsection{Mappings and Dataflows}




Each layer of a DNN can be expressed as a multi-dimensional loop nest 
over the input and weight tensors.
A CONV2D layer has 6 loops (7 with batching) while a FC layer (i.e., GEMM operation) 
has 3 loops. At runtime, these loop bounds may be smaller than the hardware resources available (leading to under-utilization) or greater than them (leading to multiple temporal passes). Scheduling the loop involves splitting these loops into sub-partitions (tiles) and re-ordering the resulting loop nest. Different accelerators choose to execute different loop levels temporally (via serialization through the same PE) or spatially (via parallelization across distinct PEs, or clusters of PEs). The choice of parallelization may allow some input tensors to be multicasted to multiple PEs, amortizing expensive accesses. Similarly, parallelized partial-sum contributing to the same output allows for the use of efficient spatial reduction hardware. The number of tiling levels, 
loop order, and parallelism 
dimensions are collectively 
referred to as 
the accelerator's \textit{dataflow} in prior work~\cite{chen2016eyeriss, kwon2020maestro}.
The dataflow, 
with specific tile sizes at 
each level of the accelerator's 
buffer hierarchy, 
is called a \textit{mapping}~\cite{parashar2019timeloop, kwon2020maestro}. 
\autoref{fig:nvdla} shows a loop-nest example for the NVDLA mapping~\cite{nvdla}. We dissect it further in \autoref{sec:flexibility_axes}.

\vspace{-1mm}
\subsection{Map-Space Exploration (MSE)}
\label{sec:background_mapper}
A given mapping's efficiency is determined by the utilization it achieves and the total number of buffer accesses it makes---particularly to expensive off-chip memories.
Prior works have demonstrated that different DNN layers prefer different mappings~\cite{parashar2019timeloop, kwon2020maestro}. 
This has motivated research in tools called \textit{mappers} to find optimized mappings given a layer and hardware target.
The search-space of mappings that can run on an accelerator (aka \textit{map-space}) is often extremely large ($O(10^{24})$~\cite{shengchun2020gamma}), so these 
mappers rely on 
heuristics~\cite{parashar2019timeloop, marvel}, 
optimization methods~\cite{shengchun2020gamma, cosa} 
or ML-based surrogate models~\cite{mindmapping} for faster searches.

This notion of a map-space includes all possible loop transformations and tile sizes for a layer that a Turing-Complete computer could execute. One additional challenge with domain-specialized accelerators is that not all mappings are \emph{valid} given the target hardware's concrete parameterized resource constraints such as size of buffers, bandwidth, number of PEs, and so on.
Just as neural networks tend to \emph{over-fit} to their training data, many DNN accelerators employ hard-coded control and datapaths in order to increase efficiency on known workloads~\cite{chen2016eyeriss} and to keep the hardware simple, which limits the choices of dataflows/mappings that can run on it. 
These \textit{partially} flexible accelerators
actually pose stricter constraints to the loop transformations and tiling than the size of HW resources do. For example, a mapping tool targeting a $32 \times 32$ spatial array has more in common with the same array at size $128 \times 128$ with larger buffers, than it does with a variant that supports spatial reduction of sums. 

\subsection{Design-Space Exploration (DSE) Challenges for Flexible Accelerators}
\label{sec:background_dse}

Traditional DSE \cite{kwon2020maestro, parashar2019timeloop} for inflexible accelerators is similar to MSE: each point in the map-space reflects a potential design point that can be ``hardened'' into silicon that is specialized to do exactly that mapping of operations, with a particular provisioning of buffering, computation, and on-chip network bandwidth. In this inflexible scenario, the design-space can be explored using the same techniques and heuristics (or even the same tool) as the map-space.

Making the candidate accelerators themselves flexible means that each of the millions of potential candidate accelerator designs are now individually associated with a particular map-space that is derived from whatever hardware features it employs---e.g., flexible buffers, flexible address generation, flexible operand-delivery networks, etc (see Section \ref{sec:hw_cost}). 
The contents of these map-spaces can range from a few significantly different mappings, to septillions \cite{shengchun2020gamma} of configurations with subtle differences.
This means that every step of a DSE \emph{includes multiple MSE}, because the DSE needs to credit the design with its best-possible mapping for every benchmark DNN layer.

Rather than trying to solve this difficult co-optimization problem, we formalize flexibility through an abstracted taxonomy (Section \ref{sec:flexibility_understanding}) and precise quantification (Section \ref{sec:mapspace}). Together, these allow practical integration of flexibility into the DSE process.

\section{Axes of Accelerator Flexibility}
\label{sec:flexibility_understanding}




%

We define accelerator flexibility as \textit{``how many different ways can we compute this workload.''} In other words, flexibility aims to describe a device's ability to run different mappings (i.e., re-arrangements or transformations) of the same starting loop nest, as opposed to unrelated loop nests from different programs (which might be a measure of \textit{programmability}).

We follow a bottom-up approach towards understanding flexibility.
In this section, we identify axes 
of flexibility and qualitatively discuss the performance benefit and HW cost for supporting each.
In \autoref{sec:mapspace}, we identify the map-spaces exposed by each axis.
In 
\autoref{sec:methodology}, 
we discuss our mechanism to search through these flexibility-constrained map-spaces.

\subsection{Axes Taxonomy}
\label{sec:flexibility_axes}



 \rev{Informed from existing work of accelerators~\cite{maeri_web, qin2020sigma, planaria, tangram_asplos2019, dyhard_dac2018, nvdla, cloudtpu,eyev2, du2015shidiannao} and map space exploration tools (MAESTRO~\cite{maeri_web}, Timeloop~\cite{parashar2019timeloop}, TVM~\cite{autotvm}, and so on~\cite{yang2020interstellar, dave2019dmazerunner}), where intra-layer mappings are captured as loop nests, we identified and unified the knobs that these tools vary when doing map-space exploration. We distill flexibility into four axes\footnote{We only consider intra-layer mapping (which most accelerators target) and did not consider inter-layer mapping.}}.

\textbf{(1) Tile sizes (T):} 
The ability to change bounds and aspect ratios of data tiles from one or more operand tensors per level of the buffer hierarchy~\cite{buffets}.


\textbf{(2) Loop order (O):} The ability to change the loop orders iterated per tiling level.

\textbf{(3) Loop parallelization (P):} The ability to change which tensor dimensions are parallelized per tiling level. This represents the \textit{spatial} partitioning of data (i.e., across PEs).

\textbf{(4) Array shape (S):} 
The ability to change the logical shape and clustering of the accelerator's hardware resources. This determines the 
number of tiling levels 
and the maximum tile sizes 
for the tensor dimensions being mapped in parallel (i.e., spatially) 
over the accelerator array.

We refer to the four axes as flexibility TOPS (tiling, order, parallelism, shape), as a mnemonic with the well-known performance metric Tera-Operations Per Second. 
\autoref{fig:nvdla} shows an example of NVDLA mapping~\cite{nvdla}, where in the 6-order for loop of convolution, tile sizes are (16, 64, 3, 3, 4, 4), loop order is (K, C, R, S, Y, X), loop parallelization is (K, C) parallelism, and array shape is (64x16).





\insertFigureWidth{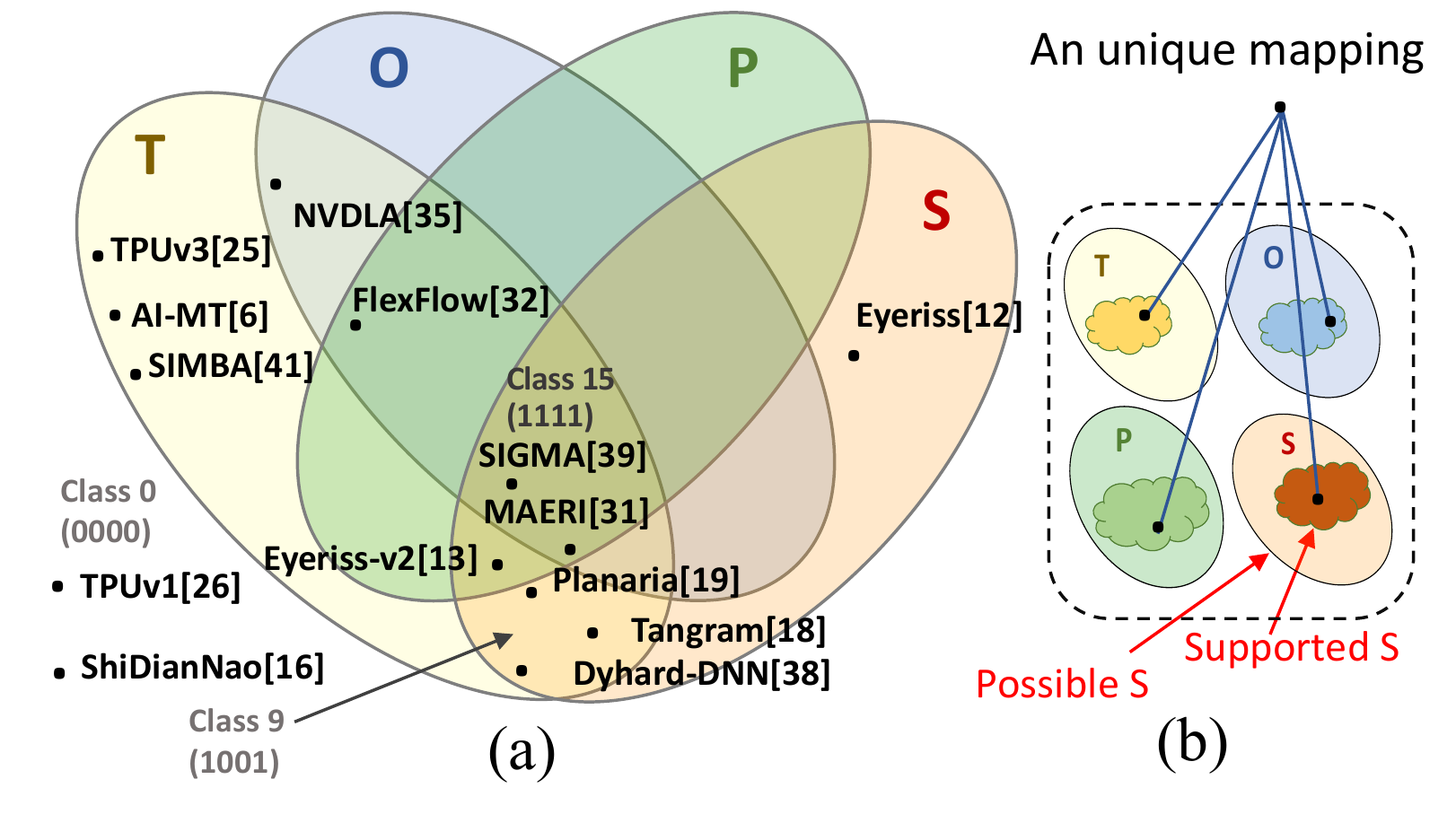}{0.75}{(a) Flexibility TOPS and prior work. (b) For any workload, an unique mapping involves choosing individual points from the map-space actually supported by the target accelerator.}

\textbf{Scope of Classification.}
Note that the four axes we described exhaustively capture the space of \textit{intra-layer mappings} which has been the target of several recent mapping optimizers~\cite{mindmapping, shengchun2020gamma, marvel, cosa, parashar2019timeloop}.
We assume that an accelerator runs each layer sequentially, consistent with state-of-the-art DNN accelerators~\cite{cloudtpu, chen2016eyeriss_jssc, simba, du2015shidiannao, nvdla, qin2020sigma, kwon2018maeri}, and flexible accelerators can change 
the mapping alone one or more of these axes for each layer.
Note that inter-layer mapping strategies (e.g., loop fusion~\cite{fusedcnn}) can be a outer-loop on top and not the focus of this study.
Further, aspects of DNN accelerators tangentially related to flexibility that we explicitly leave to future work include sparsity and compression, bit-serial arithmetic, alternate algebraic refactorings such as Winograd \cite{winograd1980convolution} or UCNN \cite{hegde2018ucnn}, and processing-in-memory and near-memory architectures.

\subsection{Classes of Accelerator Flexibility}
\label{sec:flexibility_classes}

We introduce a notation to concisely describe an accelerator as a binary vector $[X_{T},X_{O},X_{P},X_{S}]$ where:

\begin{align}
\hskip\parindent&
    X =
    \begin{cases}
 &  0 \text{ if $|Mappings_{supported}| = 1$ (inflexible)}\\ 
 &   1 \text{ if $|Mappings_{supported}| > 1$ (some flexibility)}
\end{cases}
\label{formula:map_sup}
\end{align}

\noindent $|Mappings_{supported}|$ is the number of legal mappings for a given workload (i.e., DNN layer) over a given accelerator.

Our taxonomy creates 16 accelerator classes - with Class 0 (i.e., $[X_T,X_O,X_P,X_S] = [0,0,0,0]$) being the least flexible and Class 15 (i.e., $[1,1,1,1]$) being the most. The intent of this taxonomy is to
enable a coarse classification 
of accelerators and allow a systematic study 
of cost-benefit analysis.
In \autoref{fig:flexibility_venn}(a), 
we do a best-effort classification of several current accelerators into these classes. \autoref{fig:flexibility_venn}(b) then demonstrates the role of a compiler/mapper: it generates unique mappings by selecting from the subset of the map-space that is actually supported by the target accelerator~\cite{parashar2019timeloop}.

%


\subsection{Benefits of Mapping Flexibility Per Axis}


\insertWideFigure{flexibility_benefit_v3}{The benefit of having flexibility in (a) Tile, (b) Order, (c) Parallelism, and (d) Shape. Example GEMM operation of $Z_{MN} = A_{MK} \times B_{KN}$ on 4x4 PE array. The subscript $t$ indicates that these are tiles of the overall computation. The inflexible accelerator
is weight stationary (tensor $B$), using a loop ordering of K $\rightarrow$ N $\rightarrow$ M.}

\insertWideFigure{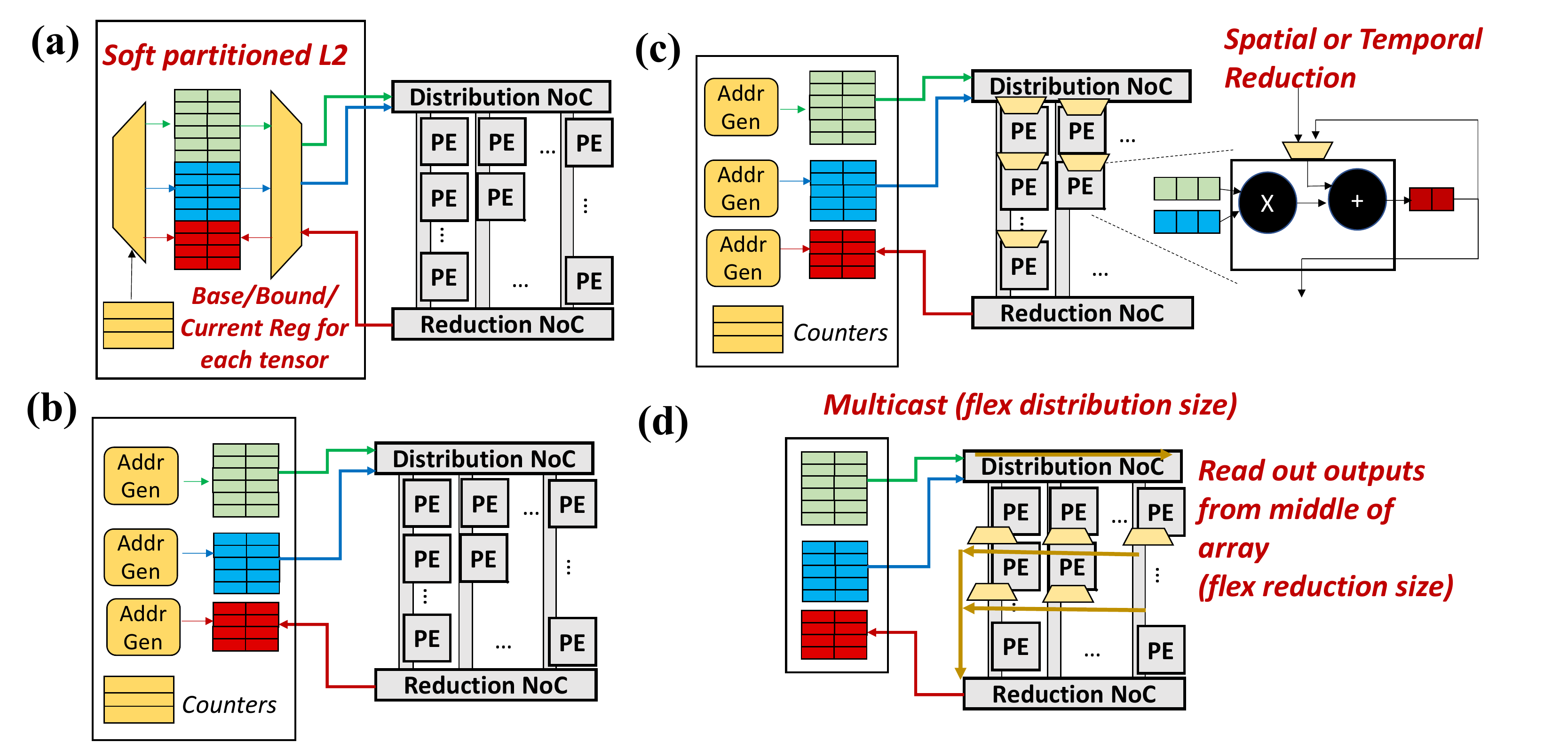}{HW Overhead for Flexibility Support (highlighted in yellow). (a) Flexible Tile Size HW Support: Soft-partitioned Scratchpads. (b) Flexible Loop Order HW Support: Configurable Address Generators. (c) Flexible Parallelism HW Support: Spatial and Temporal Reduction. (d) Flexible Shape HW Support: NoC support for PE clustering.}

In \autoref{fig:flexibility_benefit_v3} 
we show how each axis of flexibility can enhance performance and/or energy-efficiency depending on the workload dimensions.
For 
simplicity, 
we use a GEMM mapped on a 4x4 array throughout the 
example.

In \autoref{fig:flexibility_benefit_v3}(a), suppose we have a GEMM with ($M$=8, $K$=4 and $N$=8). 
It is tiled such that the size of A tiles (i.e., input activations) is 16 ($M_t$=4,$K_t$=4), and that of B tiles (i.e., weights) is 32 ($N_t$=8,$K_t$=4) - assuming this constraint comes from the sizes of the input and weight buffers around the array.
The 32 weights will 
require two iterations or passes through the 4x4 PE array during which time 16 of the weights will remain stationary; 
the two sources of inefficiency are (i) the same inputs will be read from the global buffer and streamed during each iteration (increasing energy), and (ii) switching the stationary tile will lead to a stall.
Now consider a flexible accelerator, where the global buffer is soft-partitioned 
and can 
be configured to store a tile of 32 inputs ($M_t$=8,$K_t$=4) and 16 weights ($N_t$=4,$K_t$=4) instead.
Now, the 16 weights will remain stationary
as the 32 inputs stream. Here only one iteration (although longer) will be needed, only one L2 buffer read for each input required and no stall to switch stationary tiles, potentially reducing energy and runtime.

In \autoref{fig:flexibility_benefit_v3}(b), the A tile has size 16 while the B tile has size 32.
Keeping the B tile stationary requires two iterations and two reads of the A tile, leading to a similar issue as \autoref{fig:flexibility_benefit_v3}(a).
In contrast, if we could switch the loop order to keep 
the input tile stationary instead as the flexible accelerator does, we can finish the computation in one iteration with fewer buffer reads, enhancing performance and energy-efficiency.

In \autoref{fig:flexibility_benefit_v3}(c), suppose the GEMM is not square, but rectangular.
In the default mapping discussed so far, 
the parallelism is across the K and N dimensions which leads to a utilization of only 50\%. Here the K dimension is reduced spatially across the columns (like the TPU~\cite{cloudtpu}).
If the accelerator could support flexible parallel dimensions, then we could switch to 
MN parallelism (i.e., output-stationary dataflow), performing the reduction temporally within each PE. 
Now each PE operates on a separate output and we get 100\% utilization.

In \autoref{fig:flexibility_benefit_v3}(d), we see 
that the dimensions of the weight tile (2$\times$8) do not match those of the 4x4 array, leading to two iterations, each with 50\% utilization. Flexibility in the shape of the physical array would enable it to emulate a logical 2$\times$8 aspect ratio instead, increasing utilization to 100\%.

As shown in \autoref{fig:flexibility_benefit_v3}, different axes of flexibility provide different opportunities 
to enhance utilization and/or reuse.
In fact, in some cases more than one choice exists 
for enhancing the utilization. E.g., the benefit achieved by flexible shape 
for the tall-skinny GEMM 
in \autoref{fig:flexibility_benefit_v3}(d) could also be 
achieved by switching to MN parallelism like 
\autoref{fig:flexibility_benefit_v3}(c).
This motivates this paper's focus on dissecting the performance and energy benefits along each axis of flexibility. 


\subsection{Hardware Support for Flexibility Per Axis}
\label{sec:hw_cost}

\rev{However the flexibility is not free. It comes with an area cost, as we characterize later in \autoref{table:area_cost}. For example, a multi-casting circuit takes lager area than a uni-cast; flexible choices of inputs introduce multiplexers whose area cost grows with number of choices. We characterize the specific hardware cost of each axis of flexibility via \autoref{fig:hw_cost_pic} as follows. }


\textbf{Tile:} To support tile flexibility, the accelerator needs to be able to access different tiles in the buffer. As shown in ~\autoref{fig:hw_cost_pic}(a), this costs additional Base (memory start point), Bound (data length), and Current (the current address pointer) Registers for input, weights, and outputs buffers, which are controlling/monitoring the data flow of accessing current tiles. In addition, to be able to support more varieties of tile shapes, the accelerator should have a software-controlled soft-partitioned buffer (instead of hard-partitions across each operand) to enable more tiling opportunities, shown in ~\autoref{fig:hw_cost_pic}(a) via a multiplexer/demultiplexer.


\textbf{Order:} To support different loop orderings,
the accelerator needs to allow tiles from different tensors being stationary versus streaming. 
It needs to have 
additional address counters and address generators for inputs, weights, and outputs, as shown in \autoref{fig:hw_cost_pic}(b).
Each PE also needs a register to store the counter value it should count up to before discarding the stationary tile.

\textbf{Parallelism:} To support different parallelism dimensions, we need three additional address counters and address generators. In addition, we need a dedicated multiplexer in front of each PE to select between a spatial or temporal reduction path, as shown in \autoref{fig:hw_cost_pic}(c).

\textbf{Shape:} 
Shape flexibility requires the ability for the accelerator to mimic (i) a larger  row/column by sending the same operand across multiple rows/columns and/or (ii) a smaller row/column by breaking into multiple parts.
E.g., in \autoref{fig:flexibility_benefit_v3}(d), the flexible accelerator runs two spatial reductions of size two on the same column.
Feature (i) needs a richer distribution NoC topology to multicast the operand across multiple rows/columns.
Feature (ii) requires demuxes at the output of each PE to allow it to forward the output to the neighbor or directly to the L2 buffer 
via a reduction NoC that provides these connections, as \autoref{fig:hw_cost_pic}(d) shows. Prior work~\cite{kwon2020maestro, kwon2018maeri} has discussed implementation choices for such NoCs.

We wish to highlight that beyond the area and power cost incurred by the components described above (which we find in our evaluations to not be very significant over the MAC, buffer, and NoC required by an inflexible accelerator), a key cost for adding flexibility support is that of \textit{engineering and verification}. This is why a systematic understanding of the benefits of flexibility is valuable for DNN accelerator developers.


\section{Formal Flexibility-Constrained Map Space Definition}
\label{sec:mapspace}


\insertFigureWidth{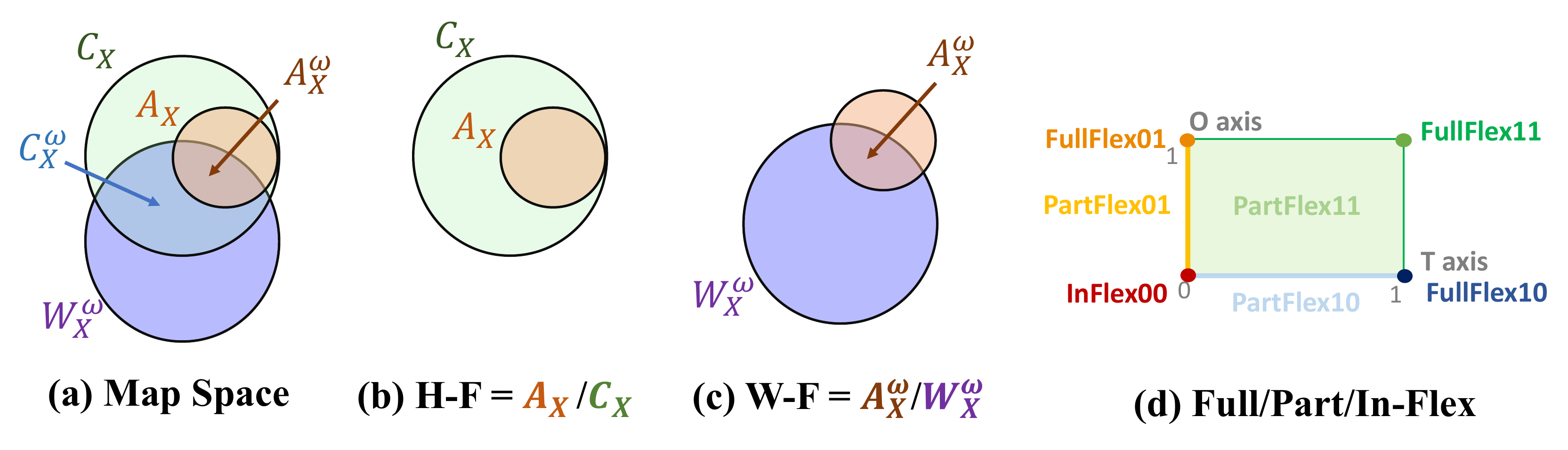}{0.85}{(a) Venn diagram of map space of a class-X target accelerator. (b) Hardware-dependent Flexion. (c) Workload-dependent Flexion. (d)A 2D example of the relationship between flexibility axes, flexibility class, and degree of flexibility. Definitions are in \autoref{table:def_table}.}

\begin{table}[t]
\mycaption{Definition of Map Space and Flexibility.}
\vspace{2mm}

\begin{tabular}{|l|l|}
\hline
$C_{X}$ & Map space of an accelerator class (class-X) \\ \hline
$A_{X}$ & Map space of the target accelerator in class-X \\ \hline
$W^{\omega}_{X}$ & Map space of workload $\omega$  \\ \hline
$C^{\omega}_{X}$ & $C_{X} \cap W^{\omega}_{X}$. Workload-dependent map space of class-X \\ \hline
$A^{\omega}_{X}$ &\makecell[l]{$A_{X} \cap W^{\omega}_{X}$. Feasible map space. Workload-dependent \\ map space of the target accelerator } \\ \hline
H-F & \makecell[l]{$A_{X}/C_{X}$. Hardware-dependent Flexion of the target \\ accelerator.  Supported map space percentage\\ in the entire class-X map space.}  \\ \hline
W-F & \makecell[l]{$A^{\omega}_{X}/W^{\omega}_{X}$. Workload-dependent Flexion of the \\ target accelerator. Supported map space percentage \\in the entire workload $\omega$ space.} \\ \hline
\end{tabular}
\label{table:def_table}
\end{table}

Given the above insights, we now formalize how hardware flexibility and workload specialization precisely affect the feasible map space.

\subsection{Definitions of Map Space and Flexibility}
\autoref{table:def_table} summarizes the key definitions. We omit the mathematical equations in the interest of space, instead use the Venn diagrams in \autoref{fig:design_space_def} to guide the discussion.

\textbf{\textit{Target accelerator: }} The accelerator whose map-space is being evaluated.

\textbf{\textit{Workload:}} Within the scope of our evaluations, this refers to a given layer of a given DNN model, since we consider the accelerators mapping layers one by one. The venn diagrams and definitions from this section will however also hold for workloads referring to full models or fused layers.

\textbf{\textit{Mapping:}} A design-point in the map space, which precisely describes the value for T, O, P, and S.

\textbf{\textit{Class-X:}} 
We use class-X to denote the class of the target accelerator (\autoref{sec:flexibility_classes}).
Note that accelerators with different configurations of HW resources such as number of PEs, buffers, and NoCs are all called class-X accelerators as long as their supporting flexibility axes are categorized into X class.
Note that the following definitions applies to all of the 16 accelerator classes, i.e., we can use it to discuss map-space and flexibility of class-0011 with two axes enabled, class-1111 with four axes enabled, and so on.

\textbf{\textit{Map space:}} the design space of the mapping of a class-X accelerator. E.g. the map space of class-0011 will expand along the two flexible axes P and S, while T and O are fixed values across all mappings. These fixed values are often practically hard-coded at accelerator design-time for simplifying hardware design (\autoref{sec:hw_cost}).

\textbf{\textit{Hardware-dependent Map Space:}}
We define two hardware-dependent map-spaces: $C_{X}$ and $A_{X}$. 

$C_{X}$ is used to denote the map space of a class-X accelerator \textit{given a constraint in total HW resources available}. Therefore, $C_{X}$ is a constrained map space. Taking Class-1xxx accelerators as an example, the total buffer size within the accelerator will directly limit the number of possible tiling choices.
Similarly, for Class-xxx1, the total number of PEs are the constraint.
No such resource constraints exist for order and parallelism.
$C_{X}$ can be viewed as the map-space of a fully-flexible accelerator within that class.
Note that $C_{X}$ is \textit{workload-agnostic}. 
For e.g., $C_{1000}$ captures all possible tile sizes that can fit in the buffers, agnostic of tile sizes that make sense for the workload (as we discuss later in this section). Thus in \autoref{fig:design_space_def}, it captures a map-space beyond what the specific workload needs.

Given the same HW resource constraint, the target accelerator in class-X might add \textit{additional constraints}, creating its own map space $A_{X}$. $A_{X}$ will always strictly be a subset of $C_{X}$, as shown in \autoref{fig:design_space_def}.
For e.g., given a total buffer size, an accelerator with hard partitioned buffers for inputs, weights and outputs will cover a smaller map-space (i.e., $A_{X}$) than 
theoretically possible in one 
that provides full flexibility in partitioning the buffers (i.e., $C_{X}$).

\textbf{\textit{Workload Map Space:}} This denotes the possible mappings for a given workload $\omega$ regardless of underlying HW, noted as ($W^{\omega}_{X}$). 
For example, for a CONV layer in ResNet50 (CONV2\_1), with 64 output and 64 input channels, 3x3 weight kernel, and 56x56 activations, the entire workload map space is all the possible combinations of 6 parameters with values from 1 to its dimension size (e.g., 1-64 for output channel). However, not every tiles can fit into the limited on-chip buffer (whose map space is captured by $C_{X}$). Therefore, for C-1xxx class, $C_{X}$ often cannot cover the full $W^{\omega}_{X}$, as \autoref{fig:design_space_def} shows.

\textbf{\textit{Workload-dependent Map Space:}} This is the intersection of hardware-dependent map space ($C_{X}$ and $A_{X}$) and the workload map space ($W^{\omega}_{X}$), which creates $C^{\omega}_{X}$ and $A^{\omega}_{X}$ in \autoref{fig:design_space_def}. 

\textbf{\textit{Feasible Map Space:}} $A^{\omega}_{X}$, the map space of the target accelerator on a given workload, is the actual map space the mapper is allowed to explore. The size of $A^{\omega}_{X}$ is the value of $|Mappings_{supported}|$ in \autoref{formula:map_sup}.

\textbf{\textit{\rev{Flexibility Fraction or Flexion}}:} The fraction of supported map space to the possible map space.

\textbf{\textit{Hardware-dependent \rev{Flexion} (H-F):}} Accelerators in the same flexibility class can be implemented with different degree of flexibility, whose feasible  map space percentage is $A_{X}$/$C_{X}$. We call this hardware-dependent Flexion (H-F).

\textbf{\textit{Workload-dependent \rev{Flexion} (W-F):}} Different workloads will have different shape and size of the map space. To understand how much portion of the map space the target accelerator can support for the current workload, we define workload-dependent Flexion (W-F) as $A^{\omega}_{X}$/$W^{\omega}_{X}$.

\subsection{Full and Partial Flexibility}
\label{sec:part_flex}


Any accelerator in class-X can have a grey-scaled degree of flexibility support (0< flexibility degree <=1) . This is illustrated for two axes of flexibility (for simplicity) in \autoref{fig:design_space_def}(d).
For ease of referring, we use the terms FullFlex-X, PartFlex-X and InFlex-X in our evaluations in \autoref{sec:eval} and \autoref{sec:future_proof}.
For e.g., FullFlex-0001 is an accelerator that supports all possible shapes S (such as MAERI~\cite{kwon2018maeri}), while 
PartFlex-0001 is an accelerator that supports a subset of shapes S (e.g., Eyeriss~\cite{chen2016eyeriss_jssc} to simplify hardware), and InFlex-0001 supports no flexibility in shape (such as the
128x128 systolic arrray within TPU-v3~\cite{cloudtpu})\footnote{InFlex-0001 is equivalent to InFlex-0000 but we make the S-bit high for ease of comparison against the partial and fully flexible versions in that class.}.
\section{Flexibility-Aware Design-Space Exploration} 
\label{sec:methodology}

\insertFigureWidthBump{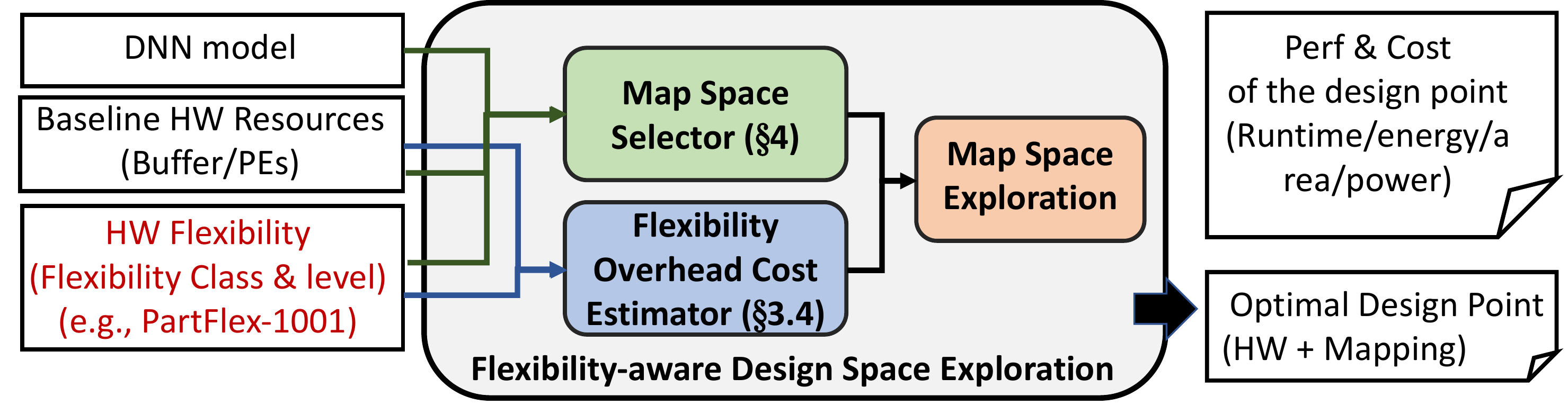}{0.8}{5pt}{Flexibility-Aware Automated Design-Space Exploration Framework.}


Ideally, the above formalism is (A) precise enough to capture a quantifiable constraint on the shape of the map-space, and (B) abstract enough not to make the DSE problem intractable.
To demonstrate this, we extend an open-source DNN mapper, GAMMA~\cite{shengchun2020gamma, gamma_web}, which uses a genetic algorithm to search through a large search-space ($\sim$10$^{24}$~\cite{shengchun2020gamma}) in a sample-efficient manner. We use the same genetic search technique for DSE and MSE where applicable.

\textbf{Modules for Flexibility-awareness.}
The native mapper only supports either inflexible accelerators (accelerators in class-0000 aka InFlex-0000) or fully flexible accelerator in class-1111 (FullFlex-1111). We extend it with two additional features:
(i) Ability to constrain the search within 16 different accelerator classes (i.e., FullFlex-xxxx). 
(ii) Ability to constrain the search further in each class for PartFlex accelerators.

These features are implemented by adding different levels of constraints to the map space of the native mapper. It includes modules for understanding the flexibility specification from input, encoding flexibility into constraints and updating the mapping exploration operators~\cite{shengchun2020gamma} correspondingly\footnote{We do not go into engineering details
in the interest of space, but will open-source the mapper upon paper acceptance.}.

\textbf{Modules for Area, Power for Estimating Cost of Flexibility.}
To estimate hardware overhead, we implemented RTL of the various components described in \autoref{fig:hw_cost_pic}, synthesized them using Synopsys DC with Nangate 15nm library~\cite{nangate-pdk} and 
used Cadence Innovus for place-and-route.
We synthesized the SRAM buffers with SAED32 education library from Synopsys, and scaled it to 15nm.
The final cost model takes in the mapping constraint file to understand the required flexibility of the accelerator and outputs its area/power/energy cost. We modularized these building block and build a cost model, which is embedded in flexibility-aware DSE (Flexibility overhead cost estimator in \autoref{fig:tool_pic_v2}).

\textbf{Toolflow.}
As shown in \autoref{fig:tool_pic_v2}, the flexibility-aware DSE takes in DNN model description, baseline HW resources, and HW flexibility specification. Three of them together defines (or so-called selects) the map space that the internal MSE tool works on. After its internal MSE tool converges or reaches a pre-defined number of sampling budget\footnote{Throughout the experiments, we set the genetic algorithm hyper-parameter as 100 populations, 100 generations, and thus having 10K sampling budget. The mutation/crossover rate and execution rate of other evolving functions as 0.5. We set these 
by applying a hyper-parameter search ~\cite{bergstra2013making}.}, it terminates the optimization and outputs the best-found design point and its HW performance (runtime, energy, area, and power).



\section{Evaluation I: Isolation Study on Flexibility Axes}
\label{sec:eval}

\begin{table}[!t]
\footnotesize

\mycaption{Hardware resources and baseline flexibility TOPS.}
\vspace{2mm}

\begin{tabular}{|l|l|}
\hline

HW Resources &\makecell[l]{Number of PEs: 1024, on-chip buffer: 100KB} \\ \hline
Mapping Config & \makecell[l]{\textbf{T}: [K:64,C:16,Y:3,X:3,R:3,S:3] tile size, \\ \textbf{O}: KCYXRS order, \textbf{P}: K-C parallel, \textbf{S}: 16x64 PE array}  \\ \hline
\end{tabular}
\label{table:baseline_dataflow}
\end{table}


\begin{table}[t]
\mycaption{Area cost of accelerators with different flexibility. }
\vspace{9pt}
\footnotesize
\begin{tabular}{|l|l|l|l|l|l|l|l|l|}
\hline
 & InFlex & T-Flex & O-Flex & P-Flex  & S-Flex & PartFlex & FullFlex \\ \hline
\makecell[l]{Area \\ $\mu m^{2}$} & \makecell[l]{736,843 \\ } & \makecell[l]{763,874 \\ (+0.004\%)} & \makecell[l]{738,458 \\ (+0.21\%)} & \makecell[l]{737,720 \\ (+0.11\%)} & \makecell[l]{737,055 \\ (+0.02\%)} & \makecell[l]{738,209 \\ (+0.19\%)} & \makecell[l]{739,576 \\ (+0.37\%)} \\ \hline
\end{tabular}
\label{table:area_cost}
\end{table}
\insertFigureWidth{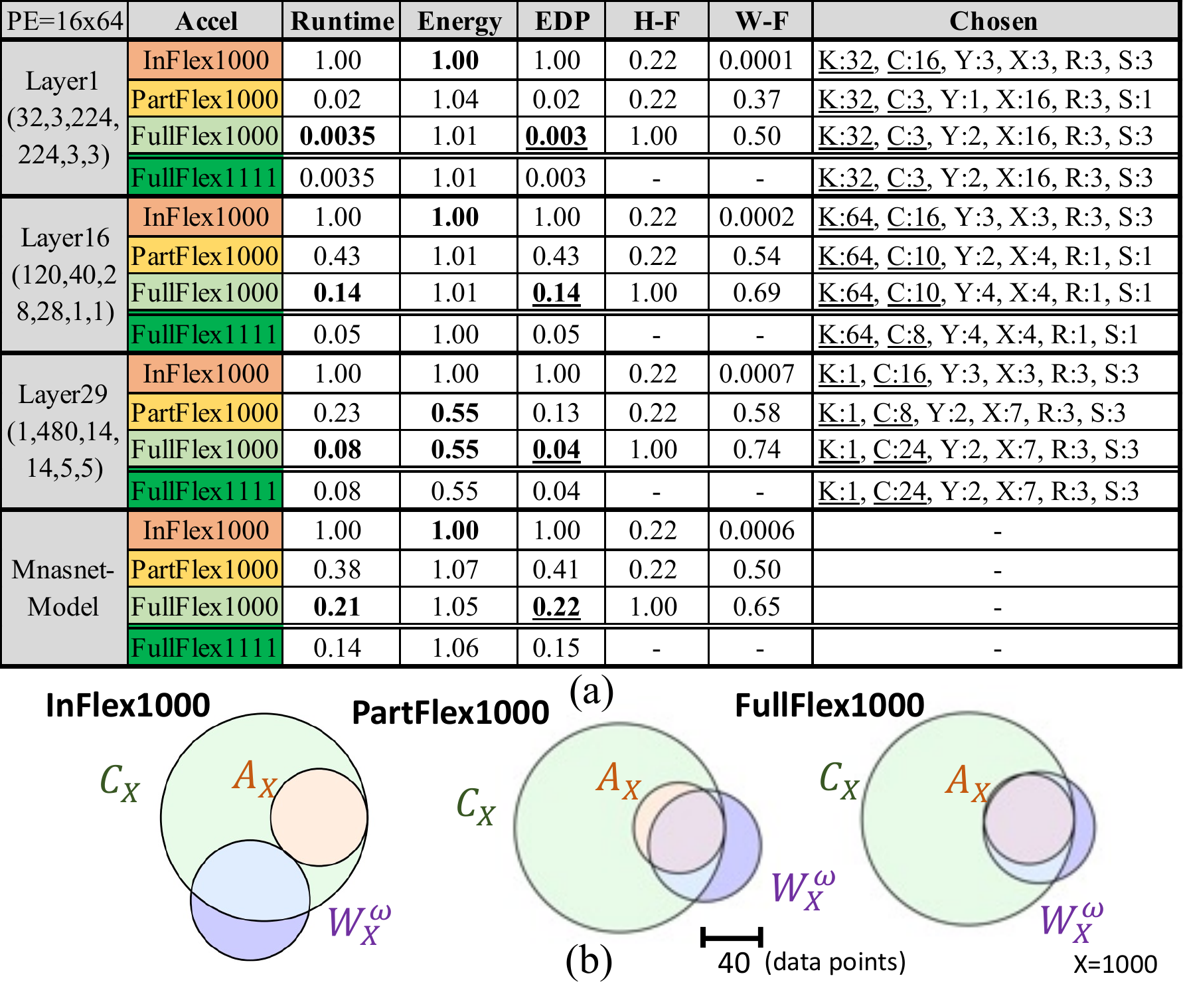}{0.75}{Tile Axis Isolation. (a) Performance comparisons of accelerators with different level of Tile flexibility running Mnasnet model. (b) Venn diagram of Tile map space on Mnasnet. We use on-chip buffer size=4K in this experiments. Scale: Total Data points in $W^\omega_T$ = $\pi(40)^2$. Runtime/Energy/EDP are normalized against values of \inflex-1000.}

\insertFigureWidth{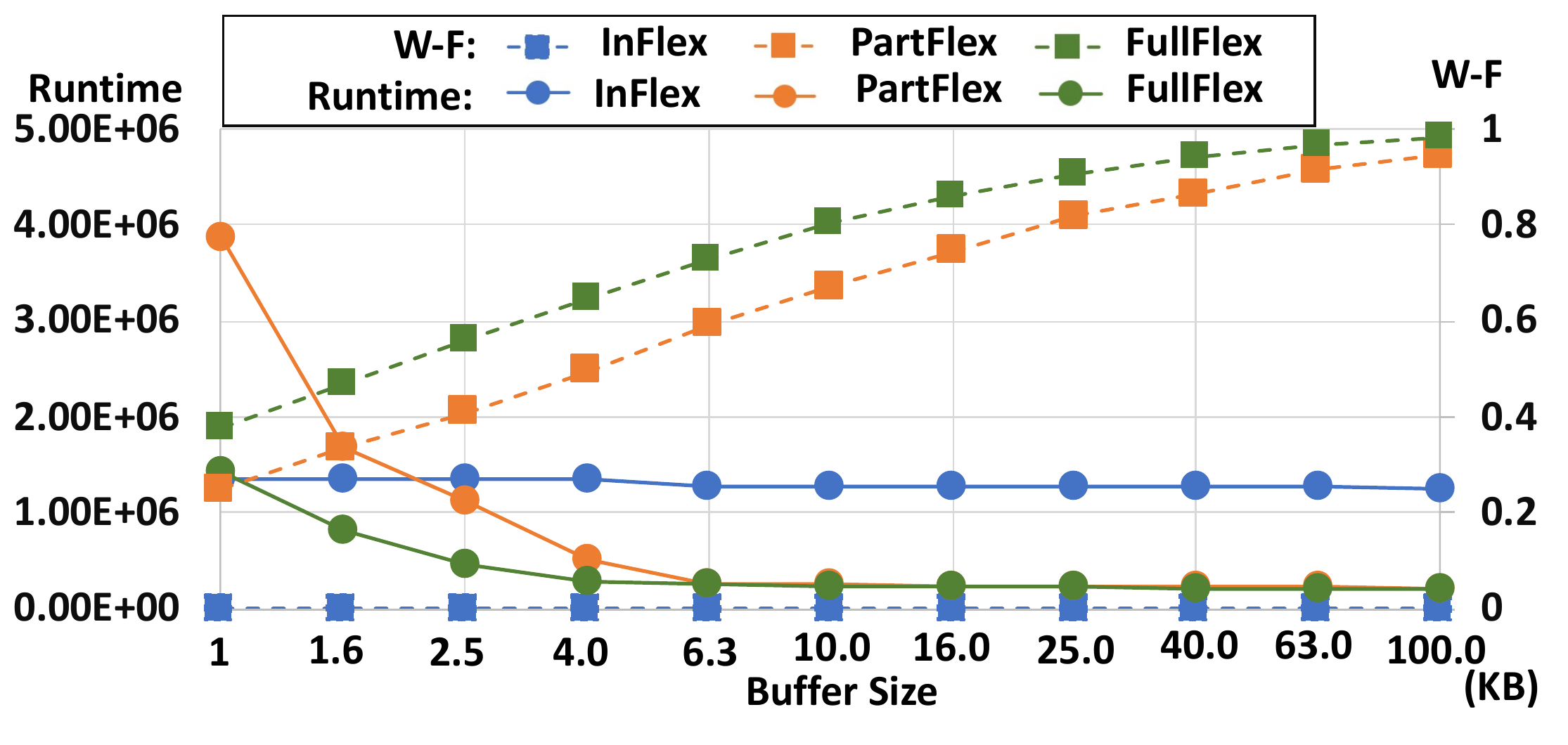}{0.7}{The runtime performance and workload Flexion of fully-Tile-flexible accelerators with different buffer sizes. (W-F: Workload-dependent Flexion.)}

\insertFigureWidth{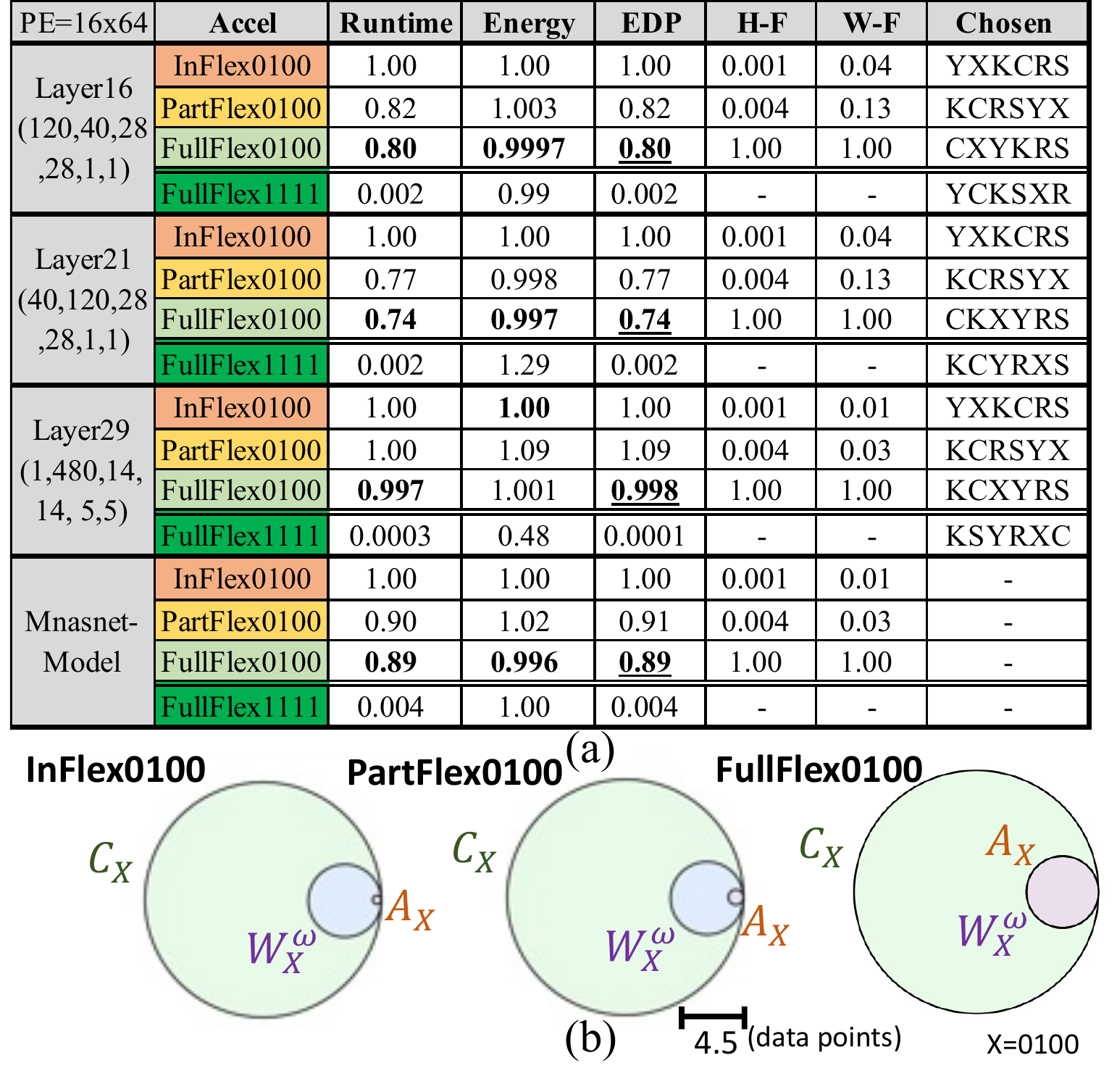}{0.75}{Order Axis Isolation. (a) Performance comparisons of accelerators with different level of Tile flexibility running Mnasnet model. (b) Venn diagram of Tile map space on Mnasnet. Runtime/Energy/EDP are normalized against values of \inflex-0100.}

\insertFigureWidth{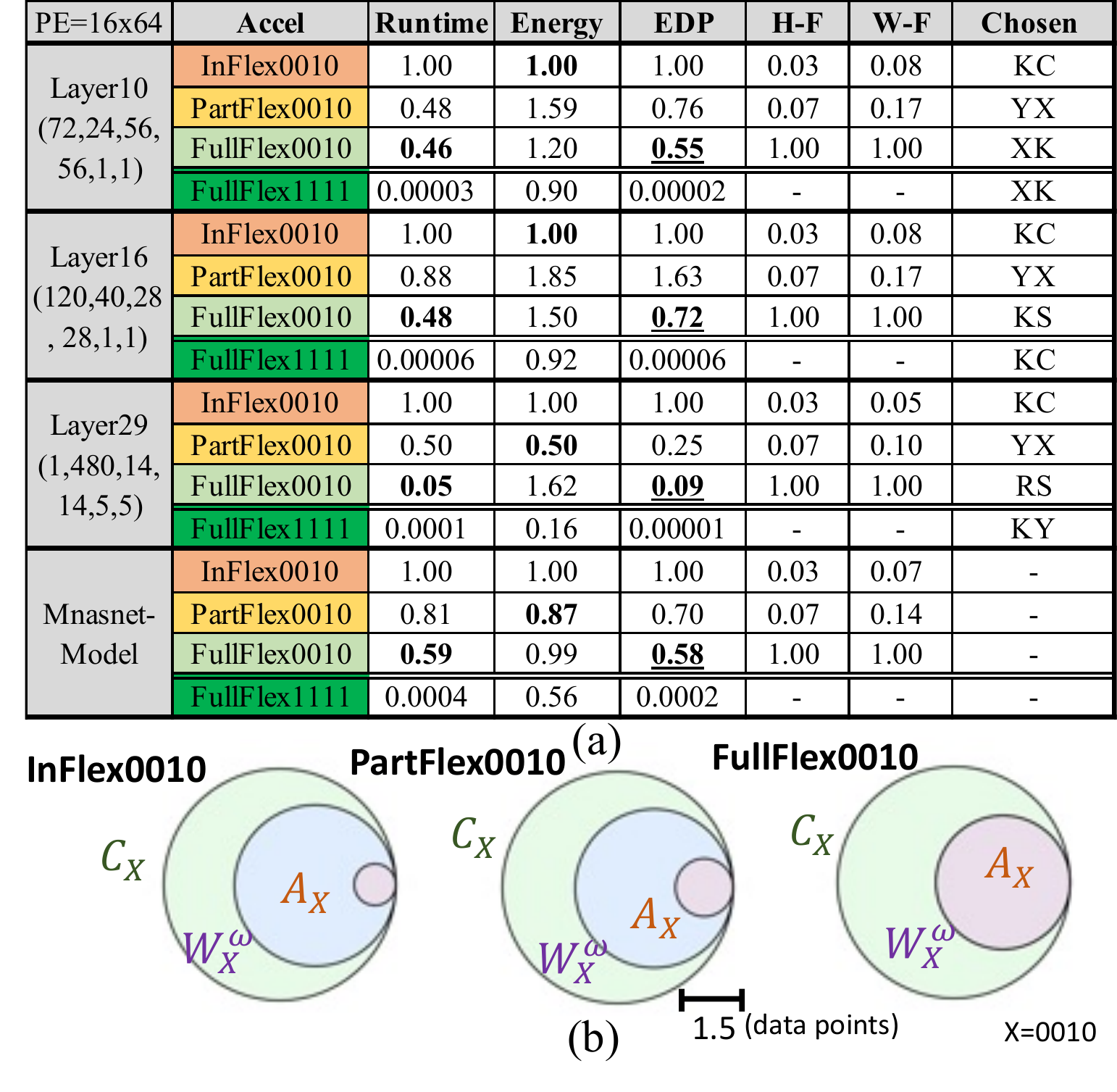}{0.75}{Parallelism Axis Isolation. Performance comparisons of accelerators with different level of Parallelism flexibility running Mnasnet model with (a) 16x64 PE array and (b) 32x32 PE array. (c) Venn diagram of Parallelism map space on Mnasnet. PRuntime/Energy/EDP are normalized against values of \inflex-0010.}

\insertFigureWidth{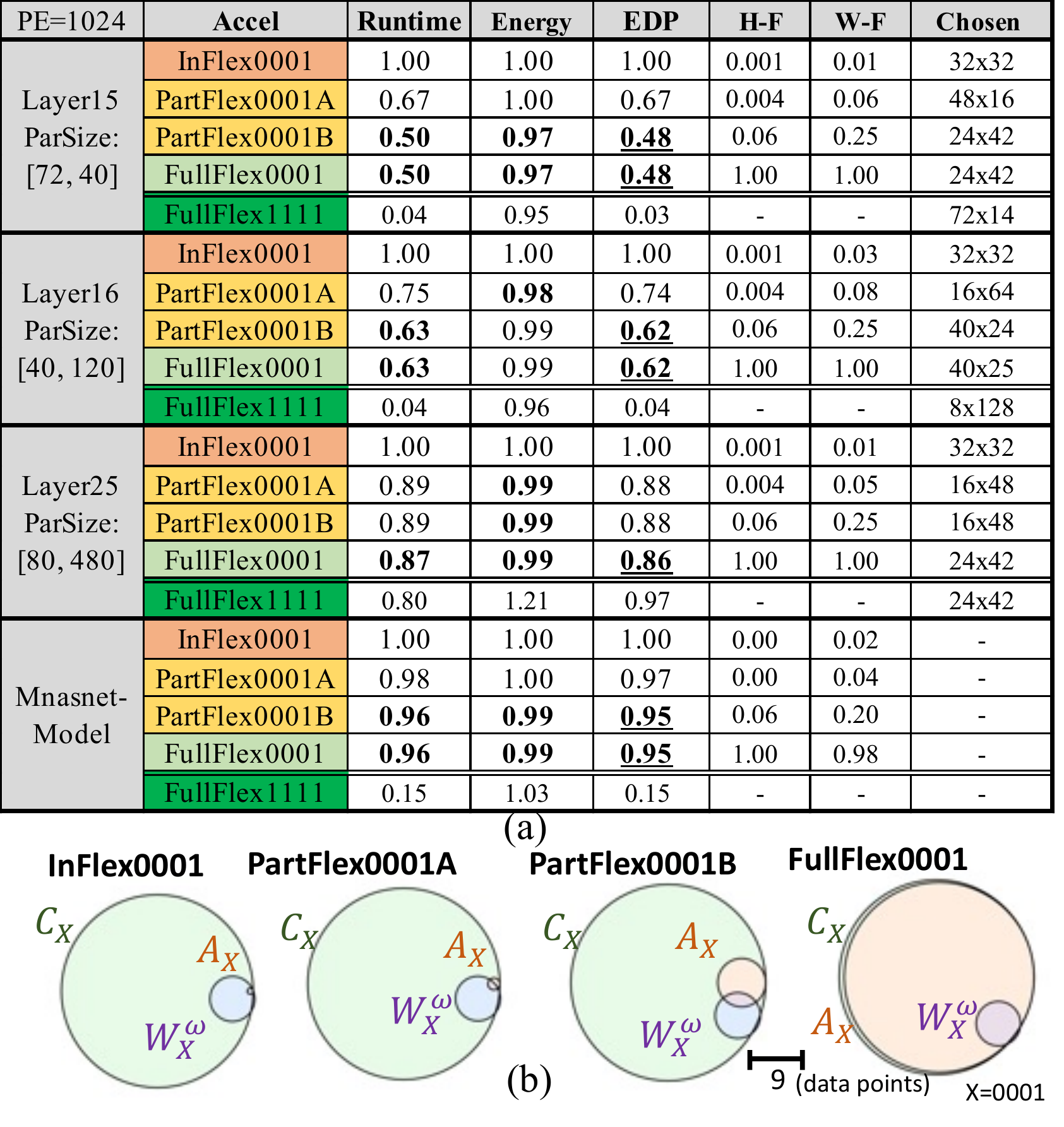}{0.75}{Shape Axis Isolation. (a) Performance comparisons of accelerators with different level of Tile flexibility running Mnasnet model. (b) Venn diagram of Tile map space on Mnasnet. Runtime/Energy/EDP are normalized against values of \inflex-0001.}

\insertFigureWidth{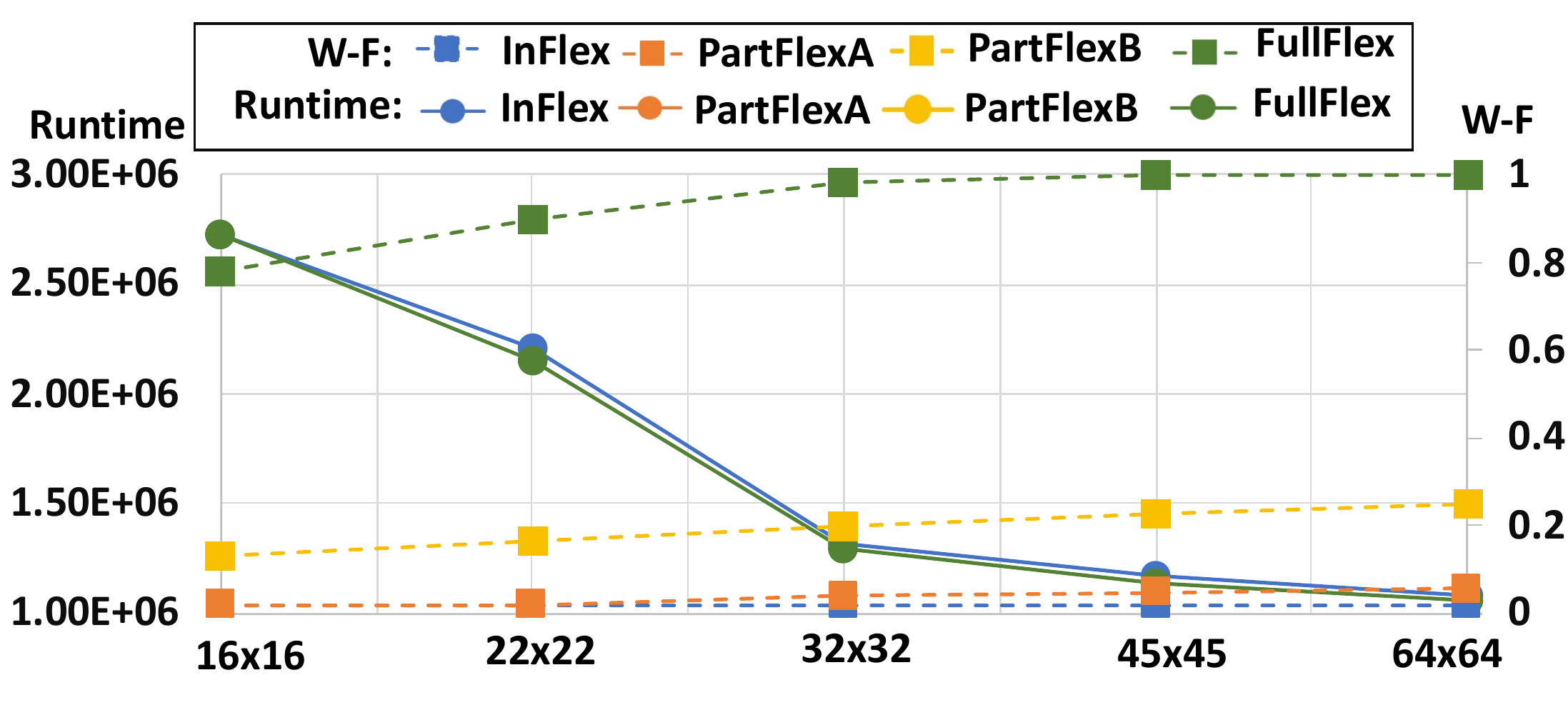}{0.7}{The runtime performance and workload Flexion of fully-Shape-flexible accelerators with different size of PE arrays. (W-F: Workload-dependent Flexion)}

\insertFigureWidthBump{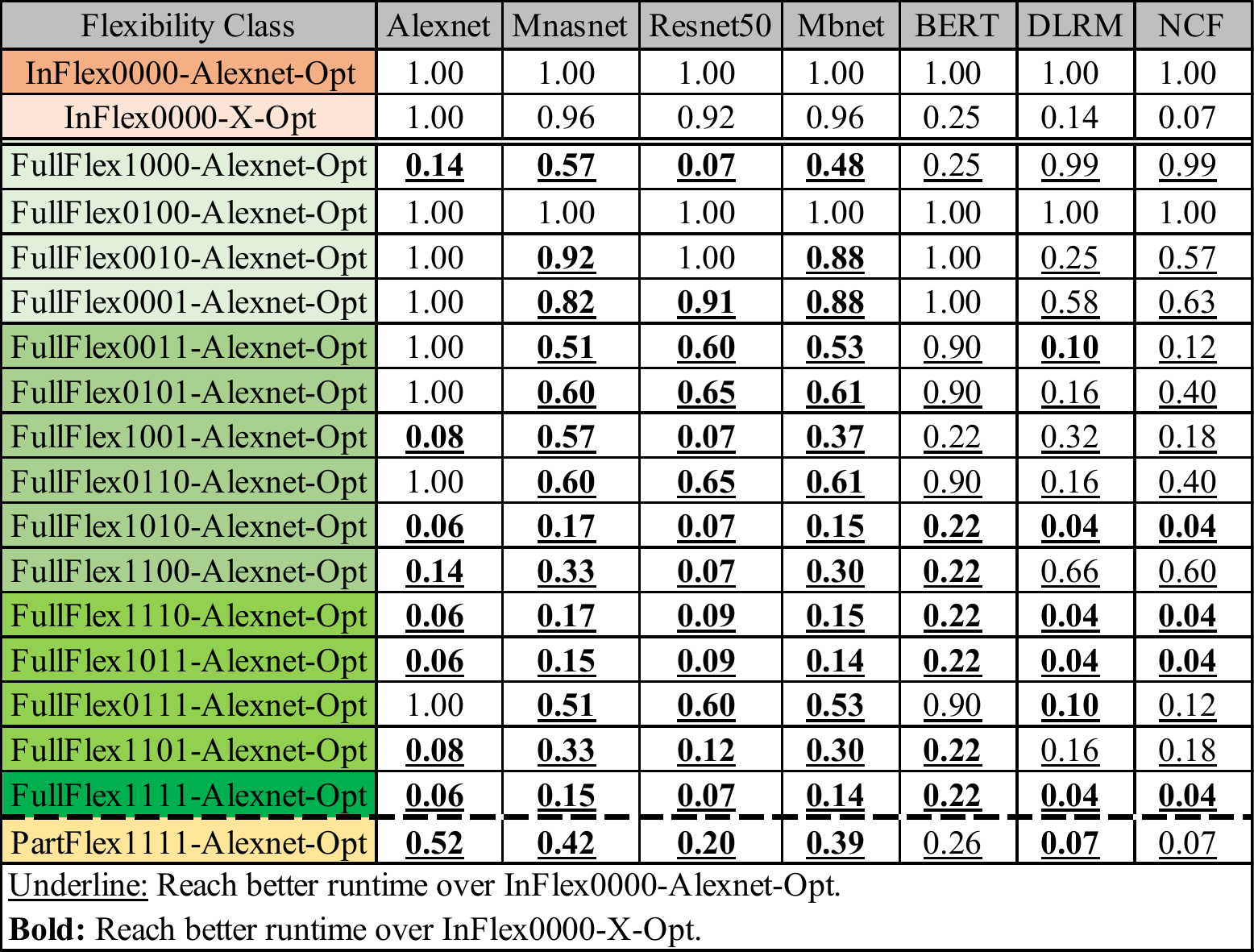}{0.75}{5pt}{Runtime performance of accelerators with different flexibility. \inflex-0000-X-Opt means class-0000 (fixed) accelerator that is optimized for DNN model X (X= Alexnet, Mnasnet, etc.). The runtime in each row is normalized by the runtime of the values of \inflex-0000-Alexnet-Opt.}

In this section, we demonstrate how architects could utilize the flexibility-aware toolflow to tractably explore different map-space constraints on candidate designs. Simultaneously, we leverage the toolflow constraints to target the DSE in order to isolate and quantify the contributions of the T/O/P/S axes.

\subsection{Methodology}

\textbf{Workloads.}
We use MnasNet~\cite{tan2019mnasnet}, which is a high speed and high accuracy model found by Neural Architecture Search (NAS), for our evaluation, since networks found by NAS have been recently achieving state-of-the-art performance.
We also look into three types of models: Vision (Alexnet~\cite{alexnet}, Resnet50~\cite{resnet}, MobilenetV2 ~\cite{sandler2018mobilenetv2}), Language  (BERT~\cite{bert}), and Recommendation (NCF~\cite{ncf}, DLRM~\cite{dlrm}).

\textbf{Target Accelerators.} For the targeted isolation studies, we formulate three accelerators with different levels of flexibility: i) an inflexible accelerator (\textbf{\inflex}), using a fixed and unchangeable TOPS configuration, ii) a partially flexible accelerator (\textbf{\partflex}) with some constrained capabilities that we describe in the relevant sections, and iii) a fully flexible accelerator (\textbf{\fullflex}), which can support the full map-space, as was described earlier in \autoref{sec:part_flex}.
These accelerators are built with the HW resources and baseline configuration listed in \autoref{table:baseline_dataflow}, and the area overhead for different flexibility is shown in \autoref{table:area_cost}. For each study, we characterized the full neural network across all layers and present the overall runtime, energy, and energy-delay product (EDP) for the best-found mappings. Furthermore, we then present detailed analysis of three layers with significantly different aspect ratios from MnasNet to illustrate the correlation between performance. 

\textbf{Optimization Objective.}
Across our experiments, we set the optimization objective to minimum runtime. However, other objectives such as energy, area, energy-delay-product (EDP), performance-per-watt, and so on, are all feasible. \rev{We can also use multi-objective formulation to optimize two or more objectives simultaneously.}

\textbf{Map-space Visualization.}
Finally, for each of the targeted isolation studies, we present a Venn diagram in the style of \autoref{fig:design_space_def} drawn to 
scale (provided in each figure) with the areas capturing the number of mappings.
Each Venn diagram plots the average of the metrics across all layers of the model.

\textbf{Axis Isolation.}
We investigate the impact of standalone flexibility of Tile (class-1000), Order (class-0100), Parallelism (class-0010) and Shape (class-0001) with the target accelerators described by varying the constraints given to the tool. This aspect of the evaluation is not intended to reflect the typical usage of the tool, but rather as a limit study that characterizes the most extreme points in the overall design space.

\subsection{Isolation: Tile Flexibility}
We study the impact of tile flexibility with three kinds of accelerators: \inflex-1000 (or \inflex-0000\footnote{\inflex-X (X=0000 to 1111) indicate the same constraint in all 16 classes.}), uses the fixed baseline tile sizes; \partflex-1000 uses hard-partitioned buffer with partition ratio of 1:1:1 for weight, input, and output buffer, and uses flexible tile sizes; \fullflex-1000 uses soft-partitioned buffer and flexible tile sizes. Note that the hardware trade-off between hard-partitioning and soft-partitioning significantly increases the energy per on-chip memory access, but can also result in fewer expensive off-chip accesses \cite{parashar2019timeloop}.

\textbf{Map Space.} Hardware-dependent Flexion (H-F) $A_{X}/C_{X}$ is decided by the partitioning strategy which is around 0.22 for 1:1:1 partition. Value 0.22 is the area ratio of $A_{X}$ over $C_{X}$ in \autoref{fig:exp_T}(b). $A^{\omega}_{X}$ is the intersection of $W^{\omega}_{X}$ and $A_{X}$, showing how many different tiles that workload $\omega$ can explore and are supported by the accelerators.  $A^{\omega}_{X}$ of \inflex-1000 is 1, since it leverages only 1 choice, while $A^{\omega}_{X}$ of \partflex-1000 will be smaller than the one for \fullflex-1000, since hard-partitioning is a stricter constraint. These can be observed by the increase of overlapped area of $W^{\omega}_{X}$ and $A_{X}$ from \inflex-1000 to \fullflex-1000. Note that the workload-dependent Flexion (W-F) $A^{\omega}_{X}/W^{\omega}_{X}$ is directly related to the ratio of the supported map space to the entire map space, where larger values implies the given flexibility in the investigating accelerator can support the current workload better.


\textbf{Performance.} \partflex-1000 allows the tile sizes to be flexible, which reduce runtime by 98\%, as shown in Layer-1 (L1) in \autoref{fig:exp_T}(a). It chooses smaller K, C sizes and larger Y, X sizes to tailor for Layer-1. While enabling soft-partitioning (\fullflex-1000), with more tile choices, the runtime improves by another $6\times$. 
In Layer-16, the C tile sizes are chosen to divide the C dimension size perfectly. \hl{The tile configurations are chosen under the criterion of best runtime, and thus energy does not show significant difference. However, if we used least energy as criterion, different points would be picked, demonstrating a larger energy difference between \inflex-1000 and  \partflex-1000/\fullflex-1000. }  
Overall, \hl{with this criterion, in class-1000}, \fullflex-1000 can achieve speedup of $4.8\times$ and $1.9\times$ over \inflex-1000 and \fullflex-1000 in end-to-end Mnasnet.


\textbf{Sensitivity Analysis: Buffer Size.} The space of supported tile shapes is directly affected by the buffer size, which is the intersection area of $S_X$ and $W^{\omega}_{S}$ in \autoref{fig:exp_T}(b). As shown in \autoref{fig:exp_T_sweep}, with increasing buffer sizes, the workload-dependent Flexion increases and the runtime improves. We can observe that the runtime improvement saturates at a buffer size of around 6.4KB, which is sufficient for most of the layers in MnasNet. Conversely, when the buffer size is small, the flexibility to execute on soft-partitioned buffer (\fullflex-1000) becomes more valuable to achieve a good runtime.

\textbf{Takeaways.} For a model like Mnasnet 
with large diversity in layer dimensions, we find 
that tile flexibility is crucial for the accelerator 
to capture the map-space. For the most part, 
the hard-partitioned \partflex-1000 provides 
significant runtime benefits over an \inflex-1000 
but is strictly worse than \fullflex-1000. We also observe that tile flexibility is more crucial for accelerators with smaller buffer sizes.

\subsection{Isolation: Order Flexibility} 
The three types of accelerators are configured as follows: \inflex-0100, using the output-stationary order YXKCRS, \partflex-0100, with three order choices including output, input, and weight stationary order, and \fullflex-0100, with flexible order. 

\textbf{Map Space.} H-F $A_{X}/C_{X}$ is affected by the number of supported order choices, which depends on the available HW support described in \autoref{fig:hw_cost_pic}. For order, the map space does not depend on the number of compute or memory resources such as PEs and buffer sizes. Therefore, $C_{X}$ can encompasses all design points in $W^{\omega}_{X}$, as shown in \autoref{fig:exp_O}(b).

\textbf{Performance.} From Layer-16 in \autoref{fig:exp_O}(a), we observe that simply adding 3 out of 6! order choices---and consequently increasing workload Flexion by a small amount---improves runtime to a near-optimal value. This demonstrates that for accelerator design, partially supporting order flexibility may expose a better cost-performance trade-off, if the supported order choices expose distinct behaviors.
In our evaluation, we also found that many different orders will end up with the similar runtime. This observation motivates the design of \partflex--0100, a simpler and lower-cost accelerator that achieves similar performance as \fullflex--0100.
Overall, \hl{in class-0100}, full flexibility support can still achieve $1.12\times$ and $1.01\times$ speedup over the two baselines.

\textbf{Takeaways.} We find that the output stationary loop-order in \inflex-0100 works reasonably well across all layers. Adding support for three loop-orders (output, weight and input) in \partflex-0100 gets nearly the same performance as supporting all 6! loop orders in \fullflex-0100.

\subsection{Isolation: Parallelism Flexibility}
The three types of accelerators are as follows: \inflex-0010, using default K-C parallelism, \partflex-0010, with a choice of K-C or Y-X parallelism, and \fullflex-0010. 

\textbf{Map Space.} Since we investigate CONV-accel (6-dim) and consider two-way parallelism. We have $C_X$=6x5=30 different choices, while the $A_X$ for \inflex-0010 and \partflex-0010 is 1 and 2, respectively. 

\textbf{Performance.} \autoref{fig:exp_P}(a) shows that different workloads can have different optimum parallelism choices, and supporting only a subset of choices may not be sufficient to approach optimal performance (Layer-16, 19). Layer-29 is a depth-wise CONV, where there is no cross-input channel computation. \inflex-0010's restriction of parallelism only across K and C dimensions deprives it of other parallelism opportunities such as YX- or RS-parallelism, which could be a better choice for this layer. 
Overall \hl{in class-0010}, \fullflex-0010 consistently achieves around $1.6\times$ and $1.3\times$ better runtime relative to the fixed and partially flexible accelerators, respectively.

\textbf{Takeaways.} We observe the flexibility in parallelism is highly-sensitive to the shape of the physical array. We also see that non-conventional parallelism choices (such as XK or KS) are found by the mapper, suggesting that supporting full flexibility in parallelism choices is valuable.

\subsection{Isolation: Shape Flexibility} The accelerators compared are as follows: \inflex-0001, using a square PE array, \partflex-0001-A, a flexible PE array composed of a modular $16\times16$ PE-array building block, \partflex-0001-B, a flexible PE array with a $4\times4$ building block, and \fullflex-0001.

\textbf{Map Space.} $C_X$ is decided by the size of the PE array. $A_X$ for the \partflex-0001 is the number of different array shapes that can be composed by the smaller building block. A smaller building block can enable more fine-grained PE array shape exploration and thus exposes a larger configuration space, which can be observed by the different size of $A_X$ for the two partially flexible accelerators in \autoref{fig:exp_S}(b).

\textbf{Performance.} When the size of parallelism dimension is larger than the size of the building block ($32\times32$, $16\times16$, $4\times4$, or flexible), we need to fold the computation. When they are not perfectly divisible, the last fold of computation will introduce compute under-utilization, as shown in Layer-15 and Layer-16 of \partflex-0001-A in \autoref{fig:exp_S}(a). Interestingly, \partflex-0001-B can mostly reach optimum performance, since most of the layers have K, C dimension sizes that are multiples of 4. Note that we use the default K-C parallelism strategy in this experiment. \partflex-0001-B reaches almost the same performance as the fully-flexible one with only $6\%$ flexibility. 
Overall, we observe that full flexibility achieve speedup of $1.05\times$, $1.02\times$, and $1.001\times$ over the three baselines, respectively. 

\textbf{Sensitivity Analysis: Array Size.} A larger number of PEs can allow more array configuration options and hence has larger hardware Flexion, as shown in \autoref{fig:exp_S_sweep}. We can also observe the diminishing return around $45\times45$ to $64\times64$ points when the provided number of PEs is starting to become larger than the maximum K-C parallelism that some of the layers can leverage, leading to under utilization of the PEs.

\textbf{Takeaways.} We observe the flexibility in array shape is highly crucial for utilization across layers. However, it is highly sensitive to the dimensions being mapped spatially. This indicates that flexibility in P and S go hand-in-hand (\hl{more discussion on this in} \autoref{sec:future_proof}). One key observation is that a partially flexible accelerator using multiple fixed-size small arrays (e.g., 4x4) reaches almost the same performance as the fully-flexible one.
This observation is consistent with some design choices in the existing accelerator designs, e.g., the $4\times4$ matrix multiplier in CUDA tensor cores~\cite{cuda_guide}. The partially flexible accelerators are also sufficient for many manual-design CNNs, e.g., ResNet50~\cite{resnet}, VGG16~\cite{vgg}, whose dimension sizes are mostly multiples of 64 in K, C dimensions. However, shape flexibility could become increasingly important because of two use cases, as follows. (1) More and more neural network designs are relying on NAS, which could generate more diverse layer shapes. MnasNet is one such NAS network, but it is highly modularized, while more diverse networks could be developed in future to target different applications. (2) Pruning techniques can change layer shapes in diverse ways.

\section{Evaluation II: Accelerator Future-Proofing} 
\label{sec:future_proof}

In this experiment, we demonstrate how this work enables a first-of-its-kind experiment: assessing the advantage of adding flexibility to an accelerator at design time in terms of future-proofing, quantifying the trade-off between hardware cost and the performance gain on ``future'' workloads. To start, we design an inflexible accelerator tailored for Alexnet,  \inflex-0000-Alexnet-Opt, by finding a fixed configuration of TOPS to optimize its runtime. This represents an accelerator such as Eyeriss~\cite{chen2016eyeriss_jssc} which was a optimized for a fixed workload in a certain year ($\sim$2014).

\textbf{Fixed Alexnet-Opt Accel on Future Models.}
We now progress our accelerator into the future. When new neural network models (model X) are introduced. 
We design \inflex-0000-X-Opts, representing a new accelerator optimized just for this new workload---though for comparison purposes we always use roughly the same number of hardware resources as  \inflex-0000-Alexnet-Opt. 
%
First, we use the \inflex-0000-Alexnet-Opt to run the future models and compare the performance witht the future model (X) optimized \inflex-0000-X-Opt. We observe that \inflex-0000-Alexnet-Opt can achieve similar performance to \inflex-0000-X-Opt when X is a CNN model, which follows the intuition since Alexnet is also a CNN, which might share similar characteristics.
However, for fully-connected (FC) layer dominated models such as BERT, DLRM, and NCF, we see 1/0.25 = 4x to 1/0.07=14x slowdown.
This shows that a fixed configuration can hurt performance when new models start to show notably different characteristics (e.g., CONV- vs FC-dominated).

\textbf{Single-Axes Flexible Alexnet-Opt Accel on Future Models.}
Next, we go back to the ``past'' to explore the effect of adding single-axes flexibility when designing Alexnet-Opt accelerator, as shown in rows 3-6 of \autoref{fig:exp_flex}. 

We see that in MnasNet, \fullflex-1000-Alexnet-Opt is $1.00/0.57=1.75\times$ better than \inflex-0000-Alexnet-Opt. \textit{We found tile flexibility to be crucial for the three CNNs and BERT}. However, it is not as effective in DLRM and NCF. 
The reason is as follows. In this experiment, \inflex-0000-Alexnet-Opt found Y-K parallelism to be optimal for Alexnet, and this stays valid for other CNNs, too.
BERT is composed of matrix-matrix multiplications, the GEMM (M,N,K) dimensions 
got mapped to the (K\_{conv},C,Y) dimensions of the CONV accelerator respectively. The Y-K parallelism in the accelerator helped BERT as well.
However, DLRM and NCF use matrix-vector multiplications, where the K dimension 
is 1 for the vector.
Mapping them on the \inflex-0000-Alexnet-Opt accelerator leads to under-utilization along the Y-parallelism dimension of the accelerator, leading to loss in performance.
In order to run DLRM and NCF efficiently, the ability to switch to other parallelism strategies become crucial.


\textbf{Multiple-Axes Flexible Alexnet-Opt Accel on Future Models.} 
There are 10 other variants of combinatorial flexible accelerators, as shown in row 7-16 of \autoref{fig:exp_flex}. We see that (P)+(S) together (0011) work well and show significant performance gain compared to the separated counterpart (0010, 0001). (T)+(P) with only two flexibility axes enabled can achieve comparable performance with \fullflex-1111-Alexnet-Opt in multiple cases. Interestingly, we see that (T)+(O) works much worse than (T)+(P). This shows that (T), which individually works well still needs to find the right other flexibility axes to have another level of performance boost.

\textbf{Fully Flexible Alexnet-Opt Accel on Future Models.} Finally, we investigate two kinds of flexible accelerators: \fullflex-1111-Alexnet-Opt, and \partflex-1111-Alexnet-Opt. \partflex-1111-Alexnet-Opt uses the configurations described in the evaluations in \autoref{sec:eval} (using variant A for Shape). 

While more flexibility support introduces larger area cost, it is noteworthy that it brings with potential to leverage better runtime \emph{even in the original AlexNet}. This is because AlexNet itself has a lot of variation across its layers. \emph{A fixed accelerator that optimizes for the average-case across all layers may miss the potential to reach the best case for each individual layer.}
For e.g.,  \fullflex-1111-Alexnet-Opt speeds up the baseline \inflex-0000-Alexnet-Opt by 1/0.06=16.7x times in Alexnet. Furthermore,  \fullflex-1111-Alexnet-Opt constantly achieves better performance than \inflex-0000-X-Opts.



\textbf{Takeaways.}
Accelerators optimized for certain layers
(such as CONV2D)  
are more future proof, even if the layer dimensions of future models change. This is because most models show abundant parallelism across the input and output channels.
Often tile flexibility seems sufficient to 
capture most benefits. 
However, for FC-dominated layer types, flexibility in parallelism and shape 
starts reaping more benefits since there are fewer dimensions and GEMMs can be irregular.
As models with several other layer types emerge (e.g., FC/DWCONV/Attention/...), investing in flexibility is valuable. 



\vspace{-2mm}
\section{Related Work}




\textbf{Flexible DNN accelerators.}
Flexibility support in DNN accelerators is an active topic of research, with several ``flexible'' accelerators proposed over the years~\cite{lu2017flexflow,  eyev2, kwon2018maeri,qin2020sigma, planaria}. 
The taxonomy from this work 
can tease 
out the specific axes of flexibility each provided and enable quantitative comparisons. 


\textbf{Mapping Tools.}
Several mapping tools have been proposed to search through map-spaces of flexible accelerators.
\autoref{sec:background_mapper} presents the state-of-the-art. 
In this work, we extend GAMMA~\cite{shengchun2020gamma} to support flexibility-aware optimization.

\vspace{-2mm}
\section{Conclusions and Future Work} 
\label{sec:conclusion}

This work demonstrates that increased formalism of the notion of flexibility leads to concrete benefits. Namely: (A) the ability to taxonomize existing accelerators along flexibility axes, (B) the ability to precisely quantify the shape of a map-space against a theoretical upper-bound, and (C) the ability to integrate flexibility into existing DSE flows. We identify that in several cases, partial flexibility along some of the axes is sufficient for capturing an optimized mapping. Most significantly, our evaluation of a 2014-style accelerator demonstrates that a design-time investment in flexibility can lead to concrete improvements in future-proofing after deployment time.


Of course, the field of deep learning is changing incredibly rapidly. Turing Complete platforms such as GPUs and CPUs will undoubtedly continue to play a key and irreplaceable role as the main platforms that allow machine learning experts to rapidly innovate. However, we believe that it is equally important that the most successful neural network architectures can be accelerated via hardware specialization, without ASIC designers risking complete obsolescence after deployment. Ultimately, the key reason flexible mapping support can be omitted from hardware designs is not area or energy, but simply design and verification effort, as well as mapper toolchain support.  We hope that this work will ultimately lead to more research that allows architects to quantify the benefits of ``future-proofing'' with the same level of precision as present-day budgets.
\rev{Finally, while most existing DSE framework focus on finding the ``optimal design point'', this work, armed with the ability to systematically constructing map-spaces of fully/partially flexible accelerators, opens an interesting future avenue of research to identify ``regions of optimal (i.e., high-performing) design space'', leading to better explainability of the DSE algorithms and interpretability of the accelerator deign space.}

We believe
this work will 
help both compiler developers and architects systematically develop 
flexibility-aware accelerator designs.
Given that flexibility is often not added in HW to simplify engineering effort, we envision partial flexibility being ripe for future work.
In future, we hope our analysis 
can be expanded to cover even more complex aspects of data orchestration and generalized 
beyond DNNs.

\vspace{-2cm}
\section{Acknowledgment}

\rev{This work was supported by NSF Award 1909900.}

\nocite{du2015shidiannao,cloudtpu, aimt, simba, tangram_asplos2019, dyhard_dac2018, nvdla,tpuv1}
\bibliographystyle{plain}
\bibliography{main}

\end{document}